# 3D-EPI Blip-Up/Down Acquisition (BUDA) with CAIPI and Joint Hankel Structured Low-Rank Reconstruction for Rapid Distortion-Free High-Resolution T$_2$* Mapping

Zhifeng Chen[1, 2, 3, 4], Congyu Liao[5] *, Xiaozhi Cao[5], Benedikt A. Poser[6], Zhongbiao Xu[7], Wei-Ching Lo[8], Manyi Wen[9], Jaejin Cho[2, 3], Qiyuan Tian[2, 3], Yaohui Wang[10], Yanqiu Feng[1, 11] *, Ling Xia[12, 13], Wufan Chen[1], Feng Liu[14], and Berkin Bilgic[2, 3, 15]

[1] School of Biomedical Engineering, Guangdong Provincial Key Laboratory of Medical Image Processing & Guangdong Province Engineering Laboratory for Medical Imaging and Diagnostic Technology, Southern Medical University, Guangzhou, China

[2] Athinoula A. Martinos Center for Biomedical Imaging, Massachusetts General Hospital, Charlestown, MA, USA

[3] Department of Radiology, Harvard Medical School, Charlestown, MA, USA

[4] Department of Data Science and AI, Faculty of IT, Monash University, Clayton, VIC, Australia

[5] Department of Radiology, Stanford University, Stanford, CA, USA

[6] Maastricht Brain Imaging Center, Faculty of Psychology and Neuroscience, University of Maastricht, the Netherlands

[7] Department of Radiotherapy, Cancer Center, Guangdong Provincial People's Hospital & Guangdong Academy of Medical Science, Guangzhou, China

[8] Siemens Medical Solutions, Boston, MA, USA

[9] Department of Chemical Pathology, The Chinese University of Hong Kong, Hong Kong, China

[10] Division of Superconducting Magnet Science and Technology, Institute of Electrical Engineering, Chinese Academy of Sciences, Beijing, China

[11] Guangdong-Hong Kong-Macao Greater Bay Area Center for Brain Science and Brain-Inspired Intelligence & Key Laboratory of Mental Health of the Ministry of Education, Southern Medical University, Guangzhou, China

[12] Department of Biomedical Engineering, Zhejiang University, Hangzhou, China

---

* Correspondence to: Yanqiu Feng, Ph.D., School of Biomedical Engineering, Southern Medical University, Guangzhou, China. Congyu Liao, Ph.D., Department of Radiology, Stanford University, Stanford, CA, USA. E-mail: foree@163.com.

[13] Research Center for Healthcare Data Science, Zhejiang Lab, Hangzhou, China

[14] School of Information Technology and Electrical Engineering, The University of Queensland, Brisbane, QLD, Australia

[15] Harvard-MIT Division of Health Sciences and Technology, Massachusetts Institute of Technology, Cambridge, MA, USA

# Abstract

**Purpose:** This work aims to develop a novel distortion-free 3D-EPI acquisition and image reconstruction technique for fast and robust, high-resolution, whole-brain imaging as well as quantitative $T_2^*$ mapping.

**Methods:** 3D-Blip-Up and -Down Acquisition (3D-BUDA) sequence is designed for both single- and multi-echo 3D GRE-EPI imaging using multiple shots with blip-up and -down readouts to encode $B_0$ field map information. Complementary k-space coverage is achieved using controlled aliasing in parallel imaging (CAIPI) sampling across the shots. For image reconstruction, an iterative hard-thresholding algorithm is employed to minimize the cost function that combines field map information informed parallel imaging with the structured low-rank constraint for multi-shot 3D-BUDA data. Extending 3D-BUDA to multi-echo imaging permits $T_2^*$ mapping. For this, we propose constructing a joint Hankel matrix along both echo and shot dimensions to improve the reconstruction.

**Results:** Experimental results on *in vivo* multi-echo data demonstrate that, by performing joint reconstruction along with both echo and shot dimensions, reconstruction accuracy is improved compared to standard 3D-BUDA reconstruction. CAIPI sampling is further shown to enhance the image quality. For $T_2^*$ mapping, $T_2^*$ values from 3D-Joint-CAIPI-BUDA and reference multi-echo GRE are within limits of agreement as quantified by Bland-Altman analysis.

**Conclusions:** The proposed technique enables rapid 3D distortion-free high-resolution imaging and $T_2^*$ mapping. Specifically, 3D-BUDA enables 1-mm isotropic whole-brain imaging in 22 s at 3 T and 9 s on a 7 T scanner. The combination of multi-echo 3D-BUDA with CAIPI acquisition and joint reconstruction enables distortion-free whole-brain $T_2^*$ mapping in 47 s at $1.1 \times 1.1 \times 1.0$ mm$^3$ resolution.

# 1. INTRODUCTION

Quantitative MRI has recently gained increased attention in neuroscientific and clinical applications (1–3). Echo-planar imaging (EPI) is a rapid encoding technique that has played an essential role in quantitative MRI, including but not limited to functional and diffusion MRI (4,5), parameter mapping (1,6,7), and quantitative susceptibility mapping (8,9).

Despite the high acquisition speed, susceptibility-induced geometric distortions and voxel intensity pile-ups have remained the main drawbacks of EPI (7,10). To mitigate the geometric distortions present in EPI, a range of correction approaches has been developed (5,6,11–18). A common approach is to improve the acceleration factor along the phase encoding direction with the aid of parallel imaging to mitigate distortion (19–23). However, as the acceleration factor increases, the g-factor penalty and under-sampling-associated artifacts may become severe (19,22,23). To address this, multi-shot EPI acquisition can be employed (11,13,17). By exploiting multi-shot readouts, geometric distortion and sub-sampling artifacts caused by high acceleration factors can be simultaneously reduced. Although multi-shot EPI has been beneficial, it suffers from shot-to-shot phase variations which hamper the combination of the shots predominantly in diffusion MRI (11,14). Hence, shot-to-shot phase correction techniques were introduced to remove phase errors, including navigator-based (15,18,24) and self-navigated approaches (11,13,14,17). In addition, advanced low-rank approaches utilize structured low-rank in k-space (12,25) or locally low-rank priors in the image domain (13) to help remove shot-to-shot phase variations. However, these approaches mitigate, but do not completely eliminate geometric distortion and voxel pile-ups.

To eliminate distortion, the hybrid-space SENSE approach was proposed for a blip-up/down acquisition scheme to obtain shots with alternating polarities (16). This enables the estimation of a $B_0$ field map using, e.g., FSL TOPUP (26). Hybrid-space SENSE incorporates this field map information into the multi-shot SENSE model to obtain distortion-free images (16,27). More recent work combined field map-informed parallel imaging with Hankel structured low-rank constraint to mitigate remaining phase errors, ghosts, and noise in distortion-free blip-up/down (BUDA) EPI (5,6,28–30). Previous work on BUDA focused on two-dimensional or simultaneous multislice (SMS) encoded imaging (5,6,28), and did not explore three-dimensional (3D) EPI readouts (31).

In this work, our contributions include the following:

1) a novel 3D blip-up/down sequence with controlled aliasing in parallel imaging (CAIPI) strategy across shots for better k-space coverage and improved image reconstruction is proposed for single- and multi-echo, multi-shot EPI acquisition;

2) a joint image reconstruction strategy for further correcting phase errors, boosting signal-to-noise ratio (SNR), and improving the accuracy of BUDA-EPI reconstruction; and

3) rapid, distortion-free $T_2^*$ mapping from 3-echo BUDA imaging.

The rest of this paper is organized as follows: In the next section, sequence diagrams of 3D-BUDA single- and multi-echo acquisitions are provided, and hybrid-space SENSE and standard Hankel structured low-rank scheme are introduced; subsequently, the new joint BUDA image reconstruction among both echo/contrast and shot dimensions are proposed; lastly, reconstruction evaluation criteria are listed. Section 3 demonstrates the performance of the proposed scheme *in vivo*. Section 4 discusses the advantages and limitations of the proposed method; future work is also considered. The last section draws conclusions. Part of this work was published as abstracts in the proceedings of the International Society for Magnetic Resonance in Medicine, 2020 (32) and 2022 (33).

## 2. METHODS

### 2.1. 3D-BUDA Sequence

Figure 1 shows the sequence diagram of the 3D-BUDA sequence and the k-space trajectory. Multi-echo EPI data were acquired with interleaved EPI shots of reversed polarity along the phase-encoding direction, creating opposing distortions to obtain $B_0$ field maps for model-based distortion correction (16). A 3-echo navigator was included for ghost correction (not shown) (34). A Siemens product excitation pulse with optimized amplitude modulation was used to suppress ripples, with an RF duration of 2.56 ms. Using high $R_{inplane}$ acceleration per shot, three echoes with adequate echo-times for $T_2^*$ mapping could fit within a single TR.

To obtain FOV-matched coil sensitivity maps, a three-dimensional low-resolution Fast Low Angle Shot (FLASH) acquisition was included along with the 3D-BUDA sequence.

While the number of echoes can be adjusted flexibly, we used three echoes for $T_2^*$ mapping, as detailed later. Blip-up/down phase encodings were implemented for each echo (yellow and green blips in phase encoding gradient, respectively). Different shots with staggered k-space trajectories along $k_y$ and $k_z$ dimensions were acquired using additional gradients in the pre-phasing part of $G_{PE}$ and $G_z$ for each echo, respectively. For different shots with different $k_y$ shifts, we delayed the echo-train according to (35).

To reduce the g-factor penalty of the highly accelerated data and improve the conditioning of the reconstruction, different CAIPI shifts (36–38) in the $k_y$-$k_z$ dimension were also employed in different blip-up/down shots.

**2.2. Image Reconstruction**

Coil sensitivity maps were estimated using ESPIRiT (39) from the FOV-matched low-resolution 3D FLASH reference data.

*1) Hankel Low-Rank reconstruction for single-echo 3D-BUDA*

3D-BUDA reconstruction incorporates $B_0$ forward-modeling into structured low-rank reconstruction to enable distortion-free EPI, as shown in Supporting Information Figure S1:

$$\underset{I}{argmin} \sum_t \|M_t F_t E(C \odot I_t) - d_t\|_F^2, s.t. rank(\mathcal{H}(I)) = r \quad (1)$$

Where $t$ represents the shot number index, $M$ is the sampling mask, $F_t$ is the discrete Fourier transform operator, and $E$ is the $B_0$ field map generated by *TOPUP* using FSL (http://fsl.fmrib.ox.ac.uk/fsl) (26). $C$ is the coil sensitivity maps, $I$ denotes the to-be-restored images, the $\odot$ symbol denotes the Hadamard (or elementwise) product between two matrixes, and $d_t$ represents the acquired under-sampled k-space data of $t^{th}$ shot. $\mathcal{H}(I)$ represents the Hankel low-rank matrix which enforces low-rankness across different shots, and $r$ is the target rank (usually determined by the noise level, reduction factor, distortion extent, etc). Herein, the Hankel matrix was constructed by consecutively using 9×9×9 neighborhood points in k-space from each shot as a Hankel-block and then concatenating them in the column dimension. A fast iterative algorithm is used to solve this equation. The root-mean-square error of less than 0.1% between consecutive iterations was used to check for convergence.

*2) Joint Hankel Low-Rank Image Reconstruction for Multi-echo 3D-BUDA*

We further improve the performance of structured low-rank reconstruction by utilizing the echo dimension, as shown in Figure 2:

$$\underset{I}{argmin} \sum_{t,n} \left\| M_{t,n} F_{t,n} E(C \odot I_{t,n}) - d_{t,n} \right\|_F^2, s.t. rank(\mathcal{H}(I)) = \gamma \qquad (2)$$

Where *t* is the shot index, and *n* represents the echo index. Other symbols and variables are the same as the previous Eq. (1), whereas the $\mathcal{H}(I)$ here indicates the new joint Hankel matrix.

The proposed joint reconstruction constructs a Hankel matrix along both shot and echo dimensions for multi-echo 3D-BUDA data. It enforces local k-space neighborhoods, extended across all echoes/contrasts in the shot dimension, having a structured low-rankness property. This is expected to take better advantage of the similarities across multiple images along the echo dimension. The neighborhood size for a Hankel-block is now 9×9×9×n (n: echo number). Therefore, the new size of the block-Hankel matrix is $((N_{FE} - m + 1) \times (N_{PE} - m + 1) \times (N_z - m + 1)) \times (m^3 \times n)$ ($N_{FE}$: number of the sampled points in frequency encoding dimension, $N_{PE}$: number of the sampled points in phase encoding dimension, $N_z$: number of $k_z$ partitions, *m*: kernel size), which is constructed by consecutively using previous 9×9×9×n neighborhood points as a Hankel-block and then concatenating them in the column dimension. The convergence condition is the same as above.

*3) $T_2$\* mapping*

Quantitative $T_2$\* (=1/$R_2$\*) mapping was performed using variable projection (VARPRO) (40) on the reconstructed volumes with a dictionary comprising signal evolution curves of $T_2$\* from 1 to 125 ms with a step size of 1 ms, 126 to 300 ms with a step size of 3 ms. A reference multi-echo GRE sequence was used as the gold standard $T_2^*$ mapping method.

## 3. Experiments Design

*In vivo* experiments were performed under institutional approval on six healthy volunteers.

### 3.1. Single-echo 3D-BUDA for fast whole-brain imaging at 3T and 7T

Table 1. The sequence parameters for *in vivo* single-echo 3D-BUDA experiments @ 3T

|  | Single-echo 3D-BUDA @ 3T (Magnetom Trio system) | | | |
|---|---|---|---|---|
| FOV [mm$^3$] | 224×224×128 | 224×224×128 | 224×224×128 | 224×224×128 |
| Resolution [mm$^3$] | 1×1×1 | 1×1×1 | 1×1×1 | 1×1×1 |
| TR [ms] | 72 | 72 | 72 | 72 |
| TE [ms] | 36 | 36 | 36 | 36 |
| Flip Angle [°] | 16 | 16 | 16 | 16 |
| EPI Factor | 128 | 128 | 128 | 128 |
| Number of Coil Elements | 32 | 32 | 32 | 32 |
| Number of Shots ($N_s$) | 4 | 4 | 4 | 2 |
| $R_{inplane}$ | 4 | 4 | 4 | 4 |
| $R_z$ | 1 | 2 | 2 | 1 |
| CAIPI (*Y* or *N*) | N | Y, z-shift = 1 | N | N |
| BW$_x$ [Hz/pixel] | 1116 | 1116 | 1116 | 1116 |
| BW$_y$ [Hz/pixel] | 16.85 | 16.85 | 16.85 | 16.85 |
| ACS | 32×32 | 32×32 | 32×32 | 32×32 |
| $R_{effective}$ ($R_{inplane} \times R_z / N_s$) | 1 | 2 | 2 | 2 |
| Acquisition Time [s] | 41 | 22 s | 22 s | 22 s |
| Figure | Fig. 3(a) | Fig. 3(b) | Fig. 3(c) | Fig. 3(d), (e), Fig. 4 |

CAIPI (*Y* or *N*), CAIPIRINHA (Yes or No); BW, bandwidth; ACS, autocalibration scan (y × z matrix)

Data were acquired on a 3 Tesla (T) MR Scanner (Magnetom Trio System; Siemens Healthineers, Erlangen, Germany) equipped with a 32-channel head coil. The gradient performance was gradient strength of 45 mT/m and slew rate of 200 T/m/s. The imaging parameters for 4-shot and 2-shot 3D-BUDA included: $R_{inplane} = 4$ and $R_z = 1$ or 2, FOV 224×224×128 mm$^3$, TR/TE = 72/36 ms, resolution 1×1×1 mm$^3$, and flip angle 16° (Ernst angle). The effective acceleration factor is $R_{effective}$=2 and the total acquisition times (TA) are 22 s in all cases, including a FLASH low-resolution pre-scan for coil sensitivity estimation. Detailed parameters can be found in Table 1.

Additional *in vivo* data were also acquired on a 7T MR Magnetom Scanner (Siemens Healthineers, Erlangen, Germany) using a 32-channel head coil array. The gradient performance was gradient strength of 70 mT/m and slew rate of 200 T/m/s. The imaging parameters for 2-shot 3D-BUDA included: $R_{inplane}$ = 5 and $R_z$ = 1 or 2, FOV 224×224×128 mm³, TR/TE = 50/25 ms, the spatial resolution of 1×1×1 mm³, and flip angle 16°. The effective acceleration factors are $R_{effective}$ = 2.5 and 5, and TA = 16 s ($R_z$ = 1) and TA = 9 s ($R_z$ = 2), respectively. More parameters can be found in Table 2.

Table 2. The sequence parameters for *in vivo* single-echo 3D-BUDA experiments @ 7T

|  | Single-echo 3D-BUDA @ 7T (Magnetom system) | |
|---|---|---|
| FOV [mm³] | 224×224×128 | 224×224×128 |
| Resolution [mm³] | 1×1×1 | 1×1×1 |
| TR [ms] | 50 | 50 |
| TE [ms] | 25 | 25 |
| Flip Angle [°] | 16 | 16 |
| EPI Factor | 128 | 128 |
| Number of Coil Elements | 32 | 32 |
| Number of Shots ($N_s$) | 2 | 2 |
| $R_{inplane}$ | 5 | 5 |
| $R_z$ | 1 | 2 |
| $BW_x$ [Hz/pixel] | 1313 | 1313 |
| $BW_y$ [Hz/pixel] | 25.08 | 25.08 |
| ACS | 32×32 | 32×32 |
| $R_{effective}$ ($R_{inplane} \times R_z / N_s$) | 2.5 | 5 |
| Acquisition Time [s] | 16 | 9 |
| Figure | Fig. S2 | Fig. 5 |

CAIPI (*Y* or *N*), CAIPIRINHA (Yes or No); BW, bandwidth; ACS, autocalibration scan (y × z matrix)

### *3.2. Multi-echo 3D-BUDA for rapid whole-brain T2\* mapping at 3T*

For $T_2^*$ mapping, multi-echo, multi-shot *in vivo* imaging experiments were performed on the same 3T scanner with the standard 32-element head coil. Imaging parameters for 8-shot and 4-shot 3D-BUDA included: $R_{inplane}$ = 8 and $R_z$ = 1 or 2, FOV 220×220×128 mm³, TR = 86 ms, TE = {18, 43.17, 68.34} ms, and flip angle 19°, resulting in a spatial resolution of 1.1×1.1×1.0 mm³. The effective acceleration factor is $R_{effective}$ = 2, and TA = 47 s in all cases. Detailed information of parameters can be found in Table 3.

Table 3. The sequence parameters for *in vivo* multi-echo 3D-BUDA experiments @ 3T

|  | Multi-echo 3D-BUDA @ 3T (Magnetom Trio system) | | | |
|---|---|---|---|---|
| FOV [mm³] | 220×220×128 | 220×220×128 | 220×220×128 | 220×220×128 |
| Resolution [mm³] | 1.1×1.1×1 | 1.1×1.1×1 | 1.1×1.1×1 | 1.1×1.1×1 |
| TR [ms] | 86 | 86 | 86 | 86 |
| TE [ms] | {18, 43.17, 68.34} | {18, 43.17, 68.34} | {18, 43.17, 68.34} | {18, 43.17, 68.34} |
| Flip Angle [°] | 19 | 19 | 19 | 19 |
| EPI Factor | 128 | 128 | 128 | 128 |
| Number of Coil Elements | 32 | 32 | 32 | 32 |
| Number of Shots ($N_s$) | 8 | 8 | 8 | 4 |
| $R_{inplane}$ | 8 | 8 | 8 | 8 |
| $R_z$ | 1 | 2 | 2 | 1 |

| CAIPI (*Y or N*) | N | Y, z-shift =1 | N | N |
| --- | --- | --- | --- | --- |
| BW$_x$ [Hz/pixel] | 1302 | 1302 | 1302 | 1302 |
| BW$_y$ [Hz/pixel] | 42.95 | 42.95 | 42.95 | 42.95 |
| ACS | 32×32 | 32×32 | 32×32 | 32×32 |
| $R_{effective}$ ($R_{inplane} \times R_z / N_s$) | 1 | 2 | 2 | 2 |
| Acquisition Time [s] | 92 | 47 | 47 | 47 |
| Figure | - | Fig. 6(e), (f) | Fig. 6(c), (d) | Fig. 6(a), (b) |

CAIPI (*Y or N*), CAIPIRINHA (Yes or No); BW, bandwidth; ACS, autocalibration scan (y × z matrix)

To generate a reference T$_2$* map, standard multi-echo 3D-GRE imaging experiments were also performed. The relevant imaging parameters included: FOV 220×220×128 mm$^3$, TR = 86ms, TE = {6, 18, 30, 43.17, 55, 68.34} ms, and flip angle 19°, resulting in a spatial resolution of 1.1×1.1×1.0 mm$^3$. The total acquisition time was ~35 minutes.

### *3.3. Image Reconstruction*

All image reconstructions were implemented in MATLAB (R2020b; the Mathworks, Inc., Natick, MA, USA) for off-line reconstruction using a Linux (CentOS) workstation with 48-Core Intel(R) Xeon(R) Gold 6248R CPU @ 3.00GHz and 512 GB of memory.

Parallel computing was also performed for all 3D-BUDA image reconstructions in MATLAB using the *parfor* operator by employing 48 cores. In this way, the reconstruction procedure was sped up, and it takes ~575 s to reconstruct whole-brain, single-echo 3D-BUDA data.

### *3.4. Evaluation of Accuracy*

For image quality assessment, root-mean-square error (RMSE) and Structure Similarity (SSIM) (41) were used with a fully-sampled reference.

In this work, we have used a distortion-free "fully sampled" BUDA-reconstructed image as a reference ($R_{effective}$ = 1). For the single-echo experiment at 3T, the 4-shot reference was sampled at $R_{inplane} \times R_z$ = 4×1 acceleration per shot. For the multi-echo experiment, the 8-shot reference was sampled at $R_{inplane} \times R_z$ = 8×1 acceleration per shot.

For T$_2$* mapping evaluation, Bland-Altman analysis (42) was used. Several regions of interest (ROI) were manually selected. Reference T$_2$* maps were obtained from a standard 3D GRE multi-echo sequence.

### 4. RESULTS

### *4.1. 3D-BUDA GRE-EPI for fast whole-brain imaging at 3T*

Figures 3 and 4 show reconstruction results of single-echo 3D-BUDA data at 3T. In Figure 3, we compare different sampling schemes for 3D-BUDA imaging. As shown in Figure 3 (b, c, d), with matching scan time, 4-shot $R_z$ = 2 with and without CAIPI sampling generates better reconstruction accuracy than 2-shot $R_z$ = 1 case (RMSE: 3.11% without CAIPI shift and 3.04% with CAIPI shift vs. 7.22%), while the data with the CAIPI sampling has lower RMSE (3.04% vs. 3.11%) under the same acceleration factor. All the Hankel structured low-rank BUDA image reconstruction have better RMSE than Hybrid-space SENSE (8.74% > 7.22% > 3.11% > 3.04%). SSIM values were also consistent with RMSEs.

Conventional SENSE, TOPUP, hybrid-space SENSE, and the proposed 3D-BUDA results for the same 2-shot data are shown in Figure 4. The distortion correction effect can be observed (white arrows and the two dotted lines). TOPUP, hybrid-space SENSE and BUDA reconstruction can all correct distortion due to the incorporated self-estimated field map information. The results of TOPUP are not much inferior when compared to the Hybrid-space SENSE and the BUDA reconstruction. However, TOPUP results seem to suffer from blurring, ostensibly due to the incorporated image interpolation step. At the same time, BUDA can further remove the remaining noise and phase errors in hybrid-space SENSE (the green arrows in Fig. 4). The local-ROI SNR estimates in this figure also support an SNR benefit of BUDA.

Comparing the reconstruction results with the same scan time in Figure 3, it can be seen that the proposed CAIPI sampling with Hankel low-rank constraint provides the best accuracy. When comparing different sampling patterns with the same Hankel low-rank constraint reconstruction, even the distribution of $k_y$-$k_z$ acceleration is better than simple $k_y$ acceleration for the same scan time.

### *4.2. 3D-BUDA GRE-EPI for fast whole-brain imaging at 7T*

Figure 5 and Supporting Information Figure S2 demonstrate the reconstruction results at 7T with $R_{\text{inplane}}$ = 5-fold acceleration using two shots. Reconstructions using SENSE, TOPUP, hybrid-space SENSE, and BUDA are shown. The distortion correction effect can be seen from the two dotted lines. TOPUP results still suffer from some blurring when compared with hybrid-space SENSE and BUDA. Phase errors and some remaining ghosts/artifacts were further mitigated in BUDA compared to hybrid-space SENSE. In addition, the area showing phase errors in Figure 5 ($R_{\text{inplane}} \times R_z = 5 \times 2$) was restored in the 3D-BUDA reconstruction (green dotted boxes).

### *4.3. 3D-BUDA GRE-EPI for fast whole-brain $T_2^*$ mapping*

Figure 6 shows the multi-echo, multi-shot 3D-BUDA acquisition results at $R_{\text{inplane}} = 8$. The absolute difference maps were also displayed between reference and reconstruction results. Two time-matched acceleration strategies are considered: Sampling 8-shots at $R_z = 2$-fold partition acceleration and 4-shots at $R_z = 1$. Both joint multi-echo and separate reconstructions for each echo are performed.

When echoes are separately reconstructed, $R_{\text{inplane}} \times R_z = 8 \times 2$ case using 8-shot has higher reconstruction accuracy (3.90% RMSE) than $R_{\text{inplane}} \times R_z = 8 \times 1$ with 4-shots (6.14%), due to more evenly distributed $k_y$-$k_z$ sampling pattern. In the $R_{\text{inplane}} = 8$ case, having acquired 8-shot also permitted better $B_0$ estimation using interim 4-shot blip-up and 4-shot blip-down reconstructions in *Topup* (see Supporting Information Figure S3), which were "effectively" $R_{\text{inplane}} \times R_z = 2 \times 2$-fold accelerated and could be readily reconstructed using SENSE.

Joint reconstruction with CAIPI sampling across echoes/shots further improved these results, where 8-shot $R_{\text{inplane}} \times R_z = 8 \times 2$ with CAIPI shift yielded 2.25%, and 4-shot $R_{\text{inplane}} \times R_z = 8 \times 1$ had 5.98% RMSE. Thus, the proposed joint multi-echo reconstruction with CAIPI shift provided more than 2.5-fold improvement over separate reconstructions without CAIPI shift, and ~1.7-fold improvement over separate reconstructions with CAIPI shift (3.8% RMSE).

For $T_2^*$ mapping, Figure 7 shows the Bland-Altman analysis of the proposed 3D-BUDA and standard multi-contrast 3D-GRE. The Bland-Altman plots show the mean and difference for the ROIs (see squares in Fig 7(b)) of $T_2^*$ values. The Bland-Altman analysis results demonstrated that the $T_2^*$ maps estimated by the 3D-CAIPI-BUDA sampling pattern with joint reconstruction and standard multi-echo GRE are within the limits of agreement, although minor biases (-1.3 ms) exist.

### 5. DISCUSSION

In this work, we proposed a 3D blip-up and blip-down sequence with CAIPI sampling for rapid single- and/or multi-contrast high-resolution $T_2^*$-weighted image acquisition. Contributions of this work include: (1) 3D-BUDA sequence with CAIPI sampling along shot dimensions for improved image reconstruction is proposed for rapid, high-resolution, whole-brain $T_2^*$-weighted imaging; (2) extension of this to multi-echo sampling and distortion-free $T_2^*$ mapping; and (3) exploring joint Hankel regularization along the additional echo dimension in multi-echo acquisitions, thus leading to more accurate reconstruction than separate Hankel-constrained reconstructions per echo.

### 5.1. Improved Sampling Pattern in 3D-BUDA

As explored previously, CAIPI distributes the under-sampling more evenly in k-space and helps improve image quality (36–38). Herein, 8-shot, $R_z = 2$ case can generate better results than the time-matched, 4-shot, and $R_z = 1$ case (Fig 6). Due to more evenly distributed k-space sampling locations, the CAIPI sampling pattern has lower RMSE and higher accuracy than the non-CAIPI case.

To provide high-fidelity $B_0$ maps for 3D-BUDA reconstruction, initial blip-up and -down SENSE reconstruction results must be made at moderate acceleration levels. This was made possible by using 8 shots at $R_{\text{inplane}} \times R_z = 8 \times 2$. The alternative approach of 4-shot, $R_{\text{inplane}} \times R_z = 8 \times 1$ is less favorable, as can be seen in Results (Fig. 6), since each blip-up and -down reconstruction will be made at $R_{\text{inplane}} \times R_z = 4 \times 1$. As such, joint image reconstruction combined with 8-shot CAIPI acquisition with $R_z = 2$ partition acceleration was proposed for robust $T_2^*$ mapping.

### 5.2. Separate vs. Joint BUDA Reconstruction

Herein, joint reconstruction among echo dimensions was proposed to enhance the image reconstruction accuracy of multi-echo 3D-BUDA. The proposed joint multi-echo Hankel low-rank reconstruction for 3D-EPI BUDA has minor difference from the previous (joint) reconstructions for 2D (gSlider) EPI (6). In the previous joint reconstruction, the block Hankel matrix is constructed by grouping two successive echo times where one image is acquired with blip-up and the other with blip-down encoding. In the current proposal, all echoes and shots are jointly considered during the Hankel low-rank reconstruction. It stacks data from all contrasts in the shot dimension and enforces local k-space neighborhoods, now extended across all echoes/contrasts in the shot dimension, to have low-rank property during the reconstruction. Leveraging similarities between echoes helped improve the image quality of multi-echo 3D-BUDA reconstruction, leading to more accurate $T_2^*$ imaging.

### *5.3. Parameter Selection*

The target rank was empirically selected for each experiment and was different for each imaging experiment. In our single-echo 3 T experiments in Fig. 3, the target rank r is set to 1.5. In 7 T results, this parameter is set to 1.25 or even lower for higher acceleration. In multi-echo joint reconstruction, the target rank is set to 4.5.

### *5.4. Limitations and Extensions*

For T2* maps generated by multi-echo imaging, the number of echoes presents a tradeoff between imaging speed and the accuracy of T2* mapping. If the number of echoes is less than three, the imaging time will be shortened at the cost of the accuracy of T2* mapping. On the contrary, if the number of echoes increases, the scan time will be prolonged with potentially more accurate T2* mapping. For the standard T2* map calculation, all 6 TEs of the fully sampled multi-echo GRE were used in this work. Still, this protocol was not optimized but matched to the 3D-BUDA acquisition with respect to the main contrast-encoding parameters (TE/TR/flip angle).

To insert three echoes with adequate echo times for quantitative imaging, the minimum TE (first TE) needed to be made as small as possible. Thus, high in-plane acceleration with the multi-shot acquisition was exploited. Due to the faster $T_2^*$ decay at 7T, higher in-plane acceleration factors may be needed to permit $T_2^*$ mapping using 3D-BUDA.

In this work, a 7T signal boost permits partition acceleration as the intrinsic SNR loss is accounted for. However, there are still some remaining artifacts in the 7T BUDA results (see Figure 5 and Supporting Information Figure S2); these might come from physiological noise (including cardiac, respiratory motion, and the flow of cerebrospinal fluid), which adversely affected the image quality. At 3T, the impact of physiologic noise and intravoxel dephasing can also be observed, e.g., in $T_2^*$ maps presented in Fig 7.

The specific absorption rate (SAR) issue is critical in 7 T imaging experiments, which is commonly related to flip angle, slice thickness, RF duration, transmission voltage, etc (43–45). In this study, we use a low flip angle, slab-selective RF pulse, and despite employing fat saturation, SAR limitations were not observed in the 7 T experiments.

In terms of computation time, current BUDA-based image reconstruction requires initial SENSE reconstruction and field map estimation using *Topup*, which prolongs the total reconstruction time.

Currently our sequence can only acquire echo-dependent sampling data, that is, the CAIPI sampling pattern is fixed for each echo. An echo-independent acquisition strategy may be more flexible. With staggered sampling across both shot and echo dimensions, the g-factor and SNR benefit may be further improved. Further gains in image quality can be obtained using novel CAIPI sampling (46–48), virtual coil concept (49,50), and artificial sparsity-based image reconstruction (23,51,52). For instance, skipped-CAIPI sampling with increased protocol flexibility for high-resolution EPI (47) can be applied for SNR and efficiency benefit in future 3D-BUDA. Shot-selective 2D CAIPIRINHA can also be exploited to further push the limits of acceleration and resolution (46). Furthermore, a clustering-based motion correction approach can be introduced to improve motion robustness (15). Deep learning approaches can also be investigated to enhance image quality and enable higher accelerations (53–59).

## 6. CONCLUSIONS

The proposed 3D-BUDA combines multi-shot acquisition using alternating polarities with a $B_0$-informed and structured low-rank regularized reconstruction to boost SNR and eliminate distortions. This permits the acquisition of distortion-free whole-brain data with high quality in 22 s on 3T and in 9 s at 7T at 1 mm isotropic resolution on the Siemens systems available in this work. In multi-echo imaging, the proposed joint Hankel structured low-rank reconstruction along both shot and echo dimensions further improves the reconstruction. This enables distortion-free $T_2^*$ mapping with whole-brain coverage at the resolution of $1.1 \times 1.1 \times 1.0$ mm$^3$ in 47 s.

## DATA AVAILABILITY STATEMENT

All the sample data were performed according to procedures approved by the local Internal Review Board after obtaining informed suitable written consents. All the 3T data as well as the 7T data will be made available on request, via a request to the corresponding author. Two exemplar 3D-BUDA data are made available at https://drive.google.com/drive/u/0/folders/1pMeJyXtj4FwTGWdGElT0-yIuqbZ1t15B. The main image reconstruction code that supports the findings is available at https://github.com/zjuczf168/zjuczf168.

## CONFLICT OF INTEREST STATEMENT

Wei-Ching Lo is a Staff Scientist of Siemens Healthineers, USA.

## ACKNOWLEDGEMENTS

This work was supported by the National Natural Science Foundation of China (61801205), and China Postdoctoral Science Foundation under Grant 2018M633073, and the Guangdong Medical Scientific Research Foundation under Grant A2019041; and the Science and Technology Program of Guangdong (2018B030333001), and the Major Scientific Project of Zhejiang Lab (2020ND8AD01).

NIBIB grants: R01 EB032378, R01 EB019437, R03 EB031175, R01 EB028797, P41 EB030006, U01 EB026996 and U01 EB025162.

NIMH grants: R01 MH116173.

and NVIDIA GPU grants.

Zhifeng Chen acknowledges the Office of China Postdoc Council fellowship (OCPC-20190089).

# Figures

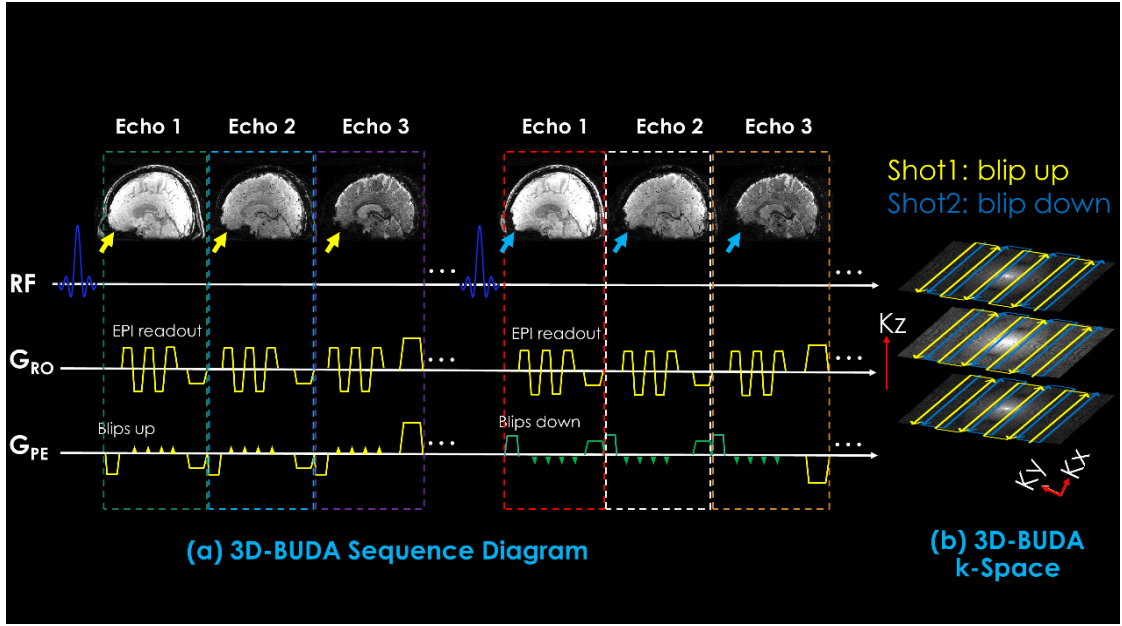

**Figure 1.** The sequence diagram of the 3D-BUDA-EPI for multi-echo imaging (a) and k-space trajectory (b). 3D slab-selective EPI data were acquired using a complementary blip-up /down acquisitions multi-shot encoding with fat saturation. Blip-up /down sampling was implemented for each echo (see yellow and green blips in phase encoding gradient), followed by a rewinder gradient before the next echo. All blip-up shots with all echoes were acquired first, then blip-down shots.

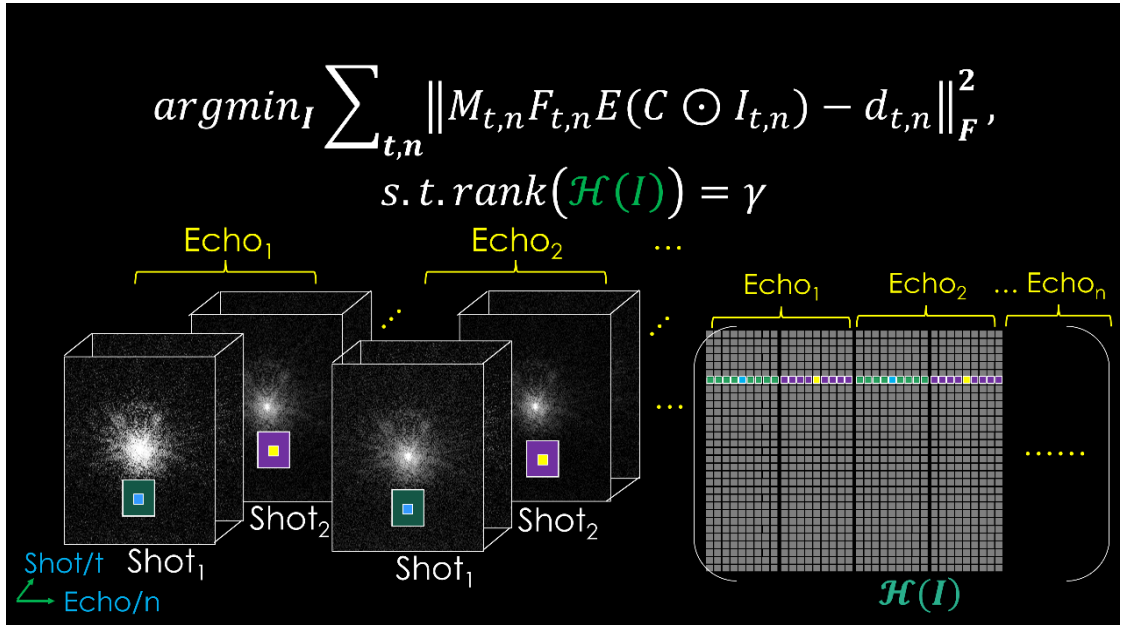

**Figure 2.** The proposed 3D-Joint-BUDA image reconstruction framework for multi-echo multi-shot GRE-EPI data. The new Hankel matrix is formed for joint image reconstruction using both echo and shot neighborhood information of the GRE-EPI dataset.

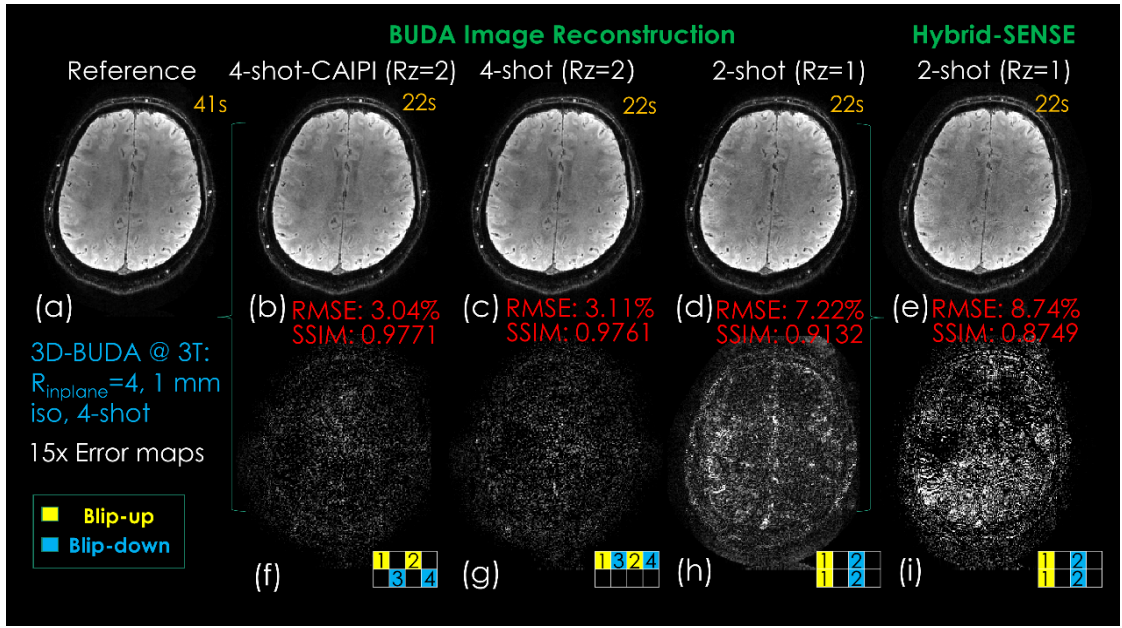

**Figure 3.** Comparison of reconstructed results and error maps of different time-matched acquisition schemes and reconstruction approaches. (a) Reference image by fully-sampled 3D-BUDA image reconstruction. (b) 4-shot $R_z = 2$ CAIPI acquisition with 3D-BUDA reconstruction result (2 blip-up shots and 2 blip-down shots). (c) 4-shot $R_z = 2$ conventional acquisition with 3D-BUDA image reconstruction (2 blip-up shots and 2 blip-down shots). (d) 2-shot $R_z = 1$ acquisition with 3D-BUDA image reconstruction (1 blip-up shot and 1 blip-down shot). (e) 2-shot $R_z = 1$ acquisition with Hybrid-space SENSE image reconstruction (1 blip-up shot and 1 blip-down shot). (f)-(i) are the corresponding difference maps. The numbers in the sampling mask represent the acquisition order of the shots.

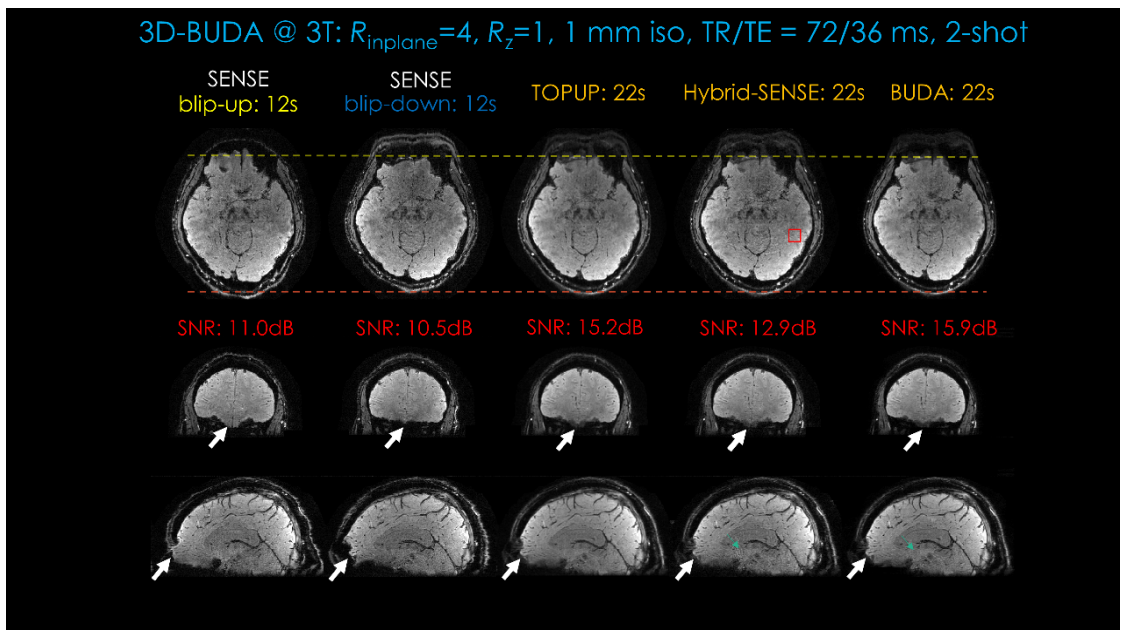

**Figure 4.** Comparison of different approaches (SENSE, TOPUP, Hybrid-space SENSE, and 3D-BUDA) on image quality and distortion-correction effect for the same 2-shot GRE-EPI BUDA dataset from 3T scanner. First column: Blip-up EPI SENSE results. Second column: Blip-down EPI SENSE results. Third column: TOPUP results. Fourth column: Hybrid-space SENSE results. Last column: The 3D-BUDA image reconstruction. Three rows are the three planes of 3D imaging. In this dataset, $R_{inplane} \times R_z$ = 4×1. The total acquisition times of blip-up EPI, blip-down EPI, TOPUP, Hybrid-space SENSE and 3D-BUDA are 12 s, 12 s, 22 s, 22 s, and 22 s, respectively. A 2-s FOV-matched FLASH low-resolution scan for coil sensitivity map is included in these experiments. Local SNR values are calculated for all the approaches. Region of interest is drawn in the Hybrid-space SENSE result in axial view (see red square in the first row).

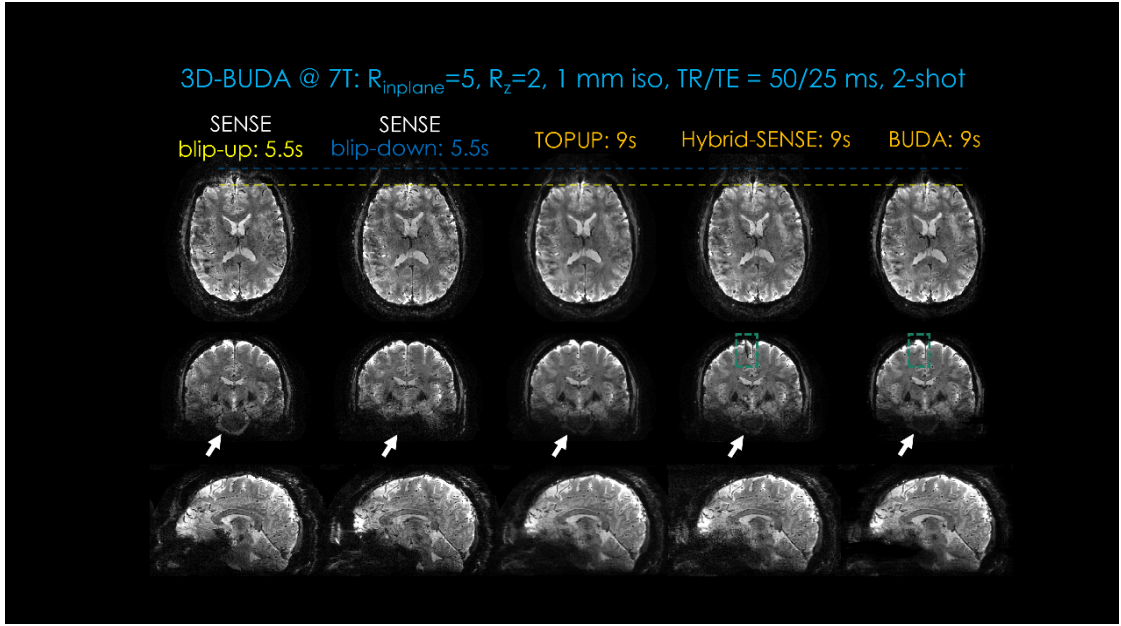

**Figure 5.** Comparison of different approaches (SENSE, TOPUP, Hybrid-space SENSE, and 3D-BUDA) on image quality and distortion-correction effect for the same 2-shot BUDA at 7T. First column: Blip-up SENSE results. Second column: Blip-down SENSE results. Third column: TOPUP results. Fourth column: Hybrid-space SENSE results. Last column: The 3D-BUDA image reconstruction results. Three rows are the three planes of 3D imaging. In this dataset, $R_{inplane} \times R_z$ = 5×2. The total acquisition times of blip-up EPI, blip-down EPI, TOPUP, Hybrid-space SENSE and 3D-BUDA are 5.5 s, 5.5 s, 9 s, 9 s, and 9 s, respectively. A 2-s FOV-matched FLASH low-resolution scan for coil sensitivity map is counted in these acquisitions.

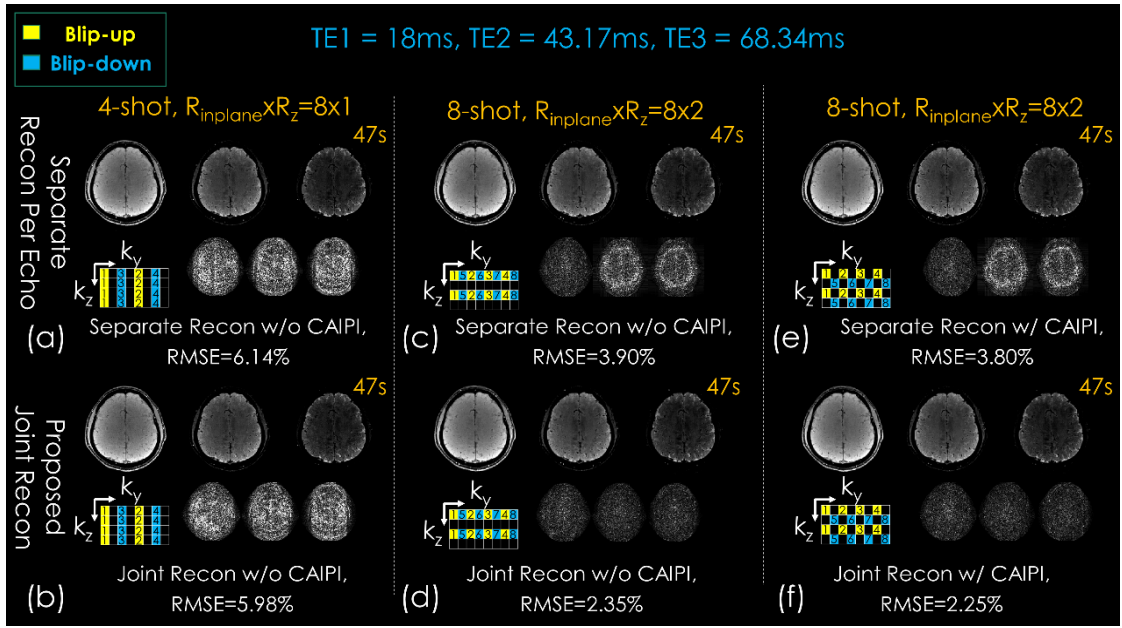

**Figure 6**. Comparison of different approaches with different sampling patterns (lower left corner of each subpart) on image quality for 3D-BUDA dataset with the same sampling amount (TA: 47 s). The difference maps were located in the lower-right corner of each subpart. First column: conventional 4-shot $R_z = 1$ without and with joint structured low-rank reconstruction. Second column: conventional 8-shot sampling without and with joint structured low-rank reconstruction. Third column: 8-shot with CAIPIRINHA sampling without and with joint structured low-rank reconstruction. Three columns in each subpart are the three echoes of 3D-BUDA imaging. The numbers in the sampling mask represent the acquisition order of the shots.

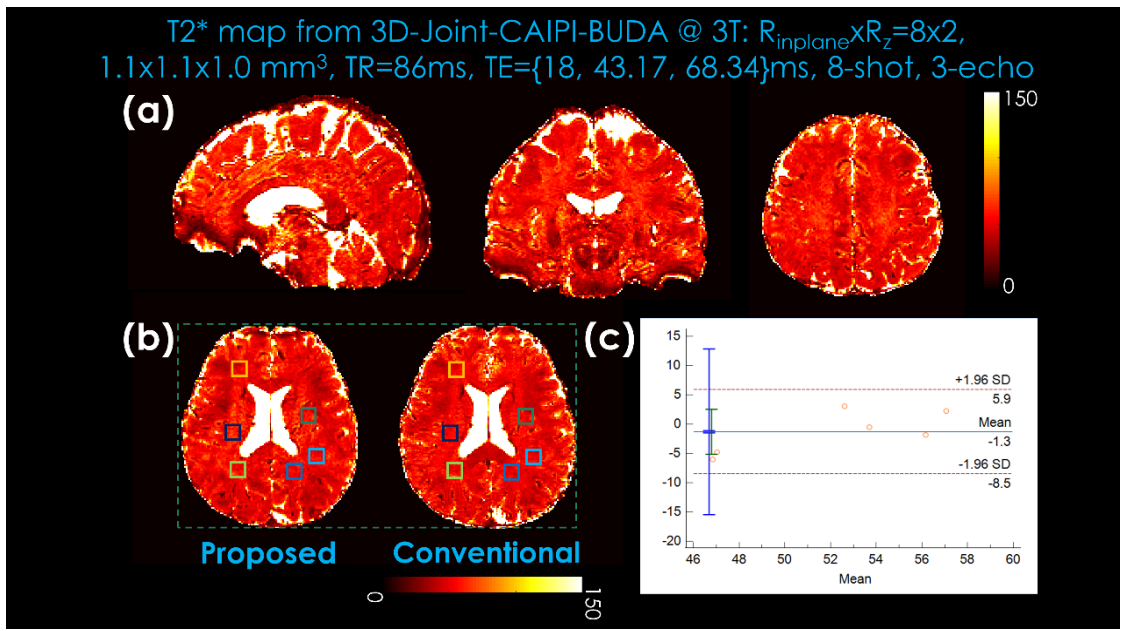

**Figure 7**. Comparison of the Bland-Altman plots displaying the mean and difference of $T_2^*$ mapping generated by 3D-Joint-CAIPI-BUDA image reconstruction and standard multi-echo GRE. (a) $T_2^*$ mapping generated by 3D-Joint-CAIPI-BUDA (8-shot, $R_{inplane} \times R_z = 8 \times 2$). (b) Selected regions of interest for Bland-Altman plots. (c) 3D-Joint-CAIPI-BUDA ($R_{inplane} \times R_z = 8 \times 2$, 8-shot) vs. standard multi-echo GRE (mean: GRE = 51.58 vs. 3D-Joint-CAIPI-BUDA = 52.88).